\RequirePackage{fix-cm}
\documentclass[smallextended]{svjour3}       
\smartqed  
\usepackage{graphicx}
\journalname{Hyperfine Interact (2013)}
\begin{document}

\title{Lorentz and CPT violation in the Standard-Model Extension
}


\author{Ralf Lehnert}


\institute{Ralf Lehnert \at
              Indiana University Center for Spacetime Symmetries, 
Bloomington, IN 47405, USA\\
              Tel.: +1-812-855-6190\\
              Fax: +1-812-855-5533 \\
              \email{ralehner@indiana.edu} 
          }

\date{Published online: 19 February 2013}

\maketitle

\begin{abstract}
Lorentz and CPT invariance are among the symmetries that can be investigated with ultrahigh precision in subatomic physics. Being spacetime symmetries, Lorentz and CPT invariance can be violated by minuscule amounts in many theoretical approaches to underlying physics that involve novel spacetime concepts, such as quantized versions of gravity. Regardless of the underlying mechanism, the low-energy effects of such violations are expected to be governed by effective field theory. This talk provides a survey of this idea and includes an overview of experimental efforts in the field.
\end{abstract}

\section{Introduction}
\label{intro}

At present, 
substantial research efforts are directed at
uncovering fundamental physics underlying 
the Standard Model (SM)
and General Relativity (GR).
While there are a number of purely theoretical approaches 
to this subject,
phenomenological progress in this field 
is inhibited by the expected Planck suppression 
of experimental signatures 
associated with these types of new physics.

Within this context, 
small departures from Lorentz and CPT invariance 
are presently being considered 
promising candidate signatures:
many theoretical ideas for underlying physics
can lead to Lorentz- and CPT-symmetry violations~\cite{lotsoftheory},
and current technology allows 
numerous types of Planck-sensitivity tests 
of these symmetries in a broad range of physical systems~\cite{ExpRev}.

For the comparison of different measurements in this field
and the identification of future tests, 
a theoretical description of small departures from Lorentz and CPT invariance 
is desirable.
A widely employed framework for this purpose, 
called the Standard-Model Extension (SME), 
is based on effective field theory~\cite{SME}. 
The SME framework 
does not represent a particular 
theoretical approach to fundamental physics 
underlying the SM and GR;
it rather describes the potential Lorentz- and CPT-violating effects 
of such theoretical ideas
at presently attainable energies 
in a general and practically model-independent way.

This work describes the subject in a top-down manner. 
Section~\ref{Theo} lists various mechanisms for Lorentz-  and CPT-symmetry breaking
in underlying physics at high energies. 
In Sec.~\ref{EFT}, 
a brief review of the main ideas behind the construction and the application 
of the SME framework is given.
Some experimental searches for deviations from Lorentz and CPT invariance 
are described in Sec.~\ref{Exp}.

\section{Mechanisms for deviations from Lorentz and CPT invariance}
\label{Theo}

Although most approaches to underlying physics 
are based on scenarios with Lorentz- and CPT-symmetric dynamics, 
they can nevertheless lead to ground states 
in which these symmetries are---at least partially---absent.
The mechanisms triggering these symmetry violations 
represent a key motivation
for Lorentz and CPT tests.
For this reason, 
we devote this section 
to an (incomplete) list of sample mechanisms for Lorentz and CPT breakdown. 
More details about the individual mechanisms 
can be found in Ref.~\cite{lotsoftheory}.

{\em Spontaneous Lorentz- and CPT-invariance violation.}---Spontaneous 
symmetry breaking (SSB) in general
is well established in physics 
and theoretically very attractive.
The basic idea behind SSB is that 
the lowest-energy state of the system, 
which is usually taken to be the vacuum,
requires the value of a suitable field to be non-vanishing.
Such non-zero vacuum expectation values (VEVs) 
often do not exhibit all dynamical symmetries of the system,
and they lead to the main observable imprints from SSB.
In the case of Lorentz and CPT breaking,
these VEVs are non-zero vector or tensor fields.  
In the flat-spacetime limit, 
these VEVs are typically taken as constant, 
which clearly shows the selection of preferred directions 
and thus the loss of exact Lorentz and CPT invariance. 
A similar, slightly more complex reasoning
also applies in gravitational contexts. 
It is difficult to fit the interactions 
required for the spontaneous violation of Lorentz and CPT invariance 
into the framework of conventional renormalizable gauge theories. 
However, 
string field theory provides a natural setting for such interactions.

{\em Spacetime-dependent scalars.}---A number of candidate theories for underlying physics
accommodate scalar fields other than the SM Higgs. 
Note also that various astrophysical measurements 
can for example be explained with new scalar fields.
In cosmological contexts, 
such scalar fields often acquire a non-vanishing global value 
with a time evolution 
driven by the expansion of the universe.
Such a background clearly violates spacetime translation invariance. 
Since translations and Lorentz transformations 
are linked in the Poincar\'e group, 
it is natural to expect that 
this breaking of translation invariance 
will also lead to a violation of Lorentz (and possibly CPT) symmetry.
This is intuitively reasonable 
because a varying scalar 
possesses a non-trivial gradient, 
and this gradient selects a preferred direction.

{\em Non-commutative field theory.}---In this approach to underlying physics, 
the basic idea is that 
coordinates are no longer real numbers. 
Instead, 
they are operators that
satisfy non-trivial commutation relations. 
For instance, 
the relation $[x^\mu,x^\nu]=i\theta^{\mu\nu}$, 
where $\theta^{\mu\nu}\neq0$ is constant, 
is typically employed as an example. 
Clearly, 
$\theta^{\mu\nu}$ is the distinguishing feature of non-commutative field theory,
so it must govern certain phenomenological effects.
But as in the above two examples, 
a constant non-dynamical $\theta^{\mu\nu}$ 
selects preferred directions in spacetime, 
so that non-commutative models 
generically violate Lorentz symmetry.

{\em Loop quantum gravity.}---Another popular approach to physics
beyond the SM and GR 
is loop quantum gravity,
which considers a version of GR that
is quantized in a particular way.
Some semiclassical analyses of loop quantum gravity 
have established various results that
are incompatible with exact Lorentz invariance. 
For instance, 
under certain reasonable physical assumptions 
both fermions and electrodynamics
receive low-energy loop-quantum-gravity corrections that
are associated with preferred directions.
As for the previous mechanisms, 
such preferred directions 
represent a violation of Lorentz symmetry.

\section{The SME test framework}
\label{EFT}

Given the multitude of sample mechanisms for Lorentz- and CPT-symmetry breaking 
listed in the previous section, 
the question arises 
how to describe {\it general} departures from these symmetries 
at low energies.
This question is exacerbated by the fact that 
for some of these mechanisms 
the direct extraction of the low-energy limit is presently unclear.
To circumvent these issues, 
it is advantageous 
to construct a suitable test framework by hand.
Such a test framework 
should be as general as possible 
while maintaining desirable physical principles. 
These ideas have led to the establishment of the SME 
mentioned in the introduction~\cite{SME}. 
What follows is a brief review of the main ingredients 
for the construction of the SME framework.

One of these ingredients is the feature 
common to all of the above sample mechanisms for Lorentz- and CPT-invariance violation:
the emergence of preferred directions in the vacuum, 
which are responsible for the symmetry breakdown.
For this reason, 
the departures from Lorentz and CPT invariance in the SME
are parametrized by non-dynamical external vector or tensor fields $b^{\mu}$, $c^{\mu\nu}$, etc.

A second ingredient is effective field theory (EFT).
This framework is widely used with great success 
in various areas of physics 
including elementary-particle,
nuclear, 
and condensed-matter physics;
it provides a general and flexible description 
of dynamical systems with 
large numbers of degrees of freedom. 
It is therefore a reasonable assumption that
the low-energy limit of potential violations of Lorentz and CPT invariance 
arising from underlying physics 
can be described within EFT.
This leads to a lagrangian formulation 
of the SME as an EFT.

A third key ingredient is coordinate independence:
although Lorentz and CPT symmetry are broken,
it should still be possible to select any suitable reference frame 
for the mathematical description of physical laws.
In other words, 
coordinates are a product of human thought
and should not acquire physical significance.
Coordinate independence is guaranteed 
if physics is formulated in terms geometrical quantities,
such as scalars, vectors, tensors, and spinors. 

These considerations lead to the following 
general structure for the SME Lagrangian ${\cal L}_{\rm SME}$:
\begin{equation}
\label{smelagr}
{\cal L}_{\rm SME}={\cal L}_{\rm SM}+{\cal L}_{\rm EH}+\delta{\cal L}_{\rm SME}\,,
\end{equation}
where ${\cal L}_{\rm SM}$ and ${\cal L}_{\rm EH}$ 
are the ordinary Standard-Model and Einstein--Hilbert contributions, 
and $\delta{\cal L}_{\rm SME}$ contains small Lorentz- and CPT-violating corrections 
constructed according the above ingredients:
\begin{equation}
\label{LVterms}
\delta{\cal L}_{\rm SME}=
-\overline{\psi}b^{\mu}\gamma_5 \gamma_\mu\psi
+i\overline{\psi}c^{\mu\nu}\gamma_\mu\partial_\nu\psi
+\ldots\,.
\end{equation}
Here, 
$b^\mu$ and $c^{\mu\nu}$ are the aforementioned examples 
for preferred directions generated by underlying physics. 
Within the context of the SME, 
they represent coefficients 
controlling the type and extent of Lorentz-symmetry breaking. 
We remark that the $b^\mu$ coefficient 
also governs certain types of CPT violation, 
while $c^{\mu\nu}$ is CPT even.
Various theoretical investigations have been performed 
within the SME~\cite{theory},
but none have found theoretical inconsistencies thus far.

In an EFT, 
one typically expects the mass-dimension three and four operators 
to dominate at low energies. 
The restriction of the SME to these operators 
is sometimes called the minimal SME, 
but higher-dimensional operators have been considered as well~\cite{HigherDim}.

\section{Experimental searches for Lorentz- and CPT-symmetry violation}
\label{Exp}

Lorentz and CPT invariance underpin 
the behavior of numerous physical systems, 
so that these symmetries can be tested 
in a correspondingly broad range of experiments~\cite{ExpRev}.
The SME test framework
discussed in the previous section 
can be used to make predictions and compare results 
for virtually all such tests.
Examples for Lorentz- and CPT-symmetry measurements
include tests with cosmic radiation~\cite{UHECR},
particle colliders~\cite{collider},
resonance cavities~\cite{cavities},
neutrinos~\cite{neutrinos}, 
and precision spectroscopy~\cite{spectroscopy}.

One might expect that 
effects from underlying physics---such as Lorentz and CPT breakdown---become
more pronounced as the energy scale is increased
because more degrees of freedom 
(including those from the underlying theory) 
can be excited. 
However, 
low-energy tests typically offer exquisite precisions that 
can more than offset 
the expected suppression of the effect.
For this reason,
low-energy physics offers excellent prospects 
in the search for 
Planck-suppressed deviations from Lorentz and CPT invariance.

Many simple low-energy systems are bound states.
A key SME prediction for bound states 
is the shift of energy levels.
Since these can be measured with
ultrahigh precision, 
they are suitable for Lorentz and CPT tests.
One class of experiments in this context 
compares the energy levels of matter and antimatter bound states. 
An example for planned measurements of this type 
is antihydrogen spectroscopy~\cite{AntiH}.
These tests would be primarily sensitive to those types 
of Lorentz breaking that
also involve CPT violation.
Another class of experiments searches for 
sidereal variations in the spacing of energy levels. 
The idea here is that 
Lorentz invariance includes rotation symmetry, 
so that there might arise effects from the rotation of the laboratory
around the Earth's axis~\cite{muonium}.
Placing the experiment on a turn table 
has also been used recently 
to search for anisotropies. 
A third class of tests seeks to measure the invariance under boosts
by studying the behavior of the energy levels 
under velocity changes. 
Many experiments in this context 
exploit the motion of the Earth around the Sun~\cite{annual}.
But space-based tests on satellites offer a viable alternative 
in this context.

Various free-particle physical systems 
can also be considered to be at comparatively low energies.
Examples in this context involve measurements 
of the muon magnetic moment at a storage ring.
Such an experiment is also affected by SME coefficients, 
and the corresponding data has previously been employed
to place constraints on Lorentz- and CPT-symmetry violation~\cite{muon}.
Another recent approach in this context 
searches for effects in the weak interaction 
by observing the decay of $^{80}$Rb or $^{20}$Na 
as a function of sidereal time.
The first results for this test are expected soon~\cite{mueller}.

\begin{acknowledgements}
The author wishes to thank the organizers 
for arranging this stimulating meeting 
and for the invitation to participate.
This work is supported by the Indiana University Center for Spacetime Symmetries. 
\end{acknowledgements}



\begin{thebibliography}{99}
%
%

\bibitem{lotsoftheory}
See, e.g., 
V.A.~Kosteleck\'y and S.~Samuel,
Phys.\ Rev.\  D {\bf 39}, 683 (1989);
J.~Alfaro, H.A.~Morales-T\'ecotl, and L.F.\ Urrutia,
Phys.\ Rev.\ Lett.\  {\bf 84}, 2318 (2000); 
S.M.~Carroll {\it et al.},
Phys.\ Rev.\ Lett.\  {\bf 87}, 141601 (2001);
J.D.~Bjorken,
Phys.\ Rev.\  D {\bf 67}, 043508 (2003);
V.A.~Kosteleck\'y {\it et al.},
Phys.\ Rev.\  D {\bf 68}, 123511 (2003);
O.~Bertolami {\it et al.},
Phys.\ Rev.\  D {\bf 69}, 083513 (2004).

\bibitem{ExpRev}
V.A.~Kosteleck\'y and N.~Russell,
Rev.\ Mod.\ Phys.\  {\bf 83}, 11 (2011).

\bibitem{SME}
D.~Colladay and V.A.~Kosteleck\'y,
Phys.\ Rev.\ D {\bf 58}, 116002 (1998);
V.A.~Kosteleck\'y and R.~Lehnert,
Phys.\ Rev.\ D {\bf 63}, 065008 (2001);
V.A.~Kosteleck\'y,
Phys.\ Rev.\ D {\bf 69}, 105009 (2004).


\bibitem{theory} 
See, e.g., 
R.~Jackiw and V.A.~Kosteleck\'y,
Phys.\ Rev.\ Lett.\  {\bf 82}, 3572 (1999);
V.A.~Kosteleck\'y {\it et al.},
Phys.\ Rev.\  D {\bf 65}, 056006 (2002);
B.~Altschul and V.A.~Kosteleck\'y,
Phys.\ Lett.\  B {\bf 628}, 106 (2005);
R.~Lehnert, 
Phys.\ Rev.\  D {\bf 68}, 085003 (2003);
J.\ Math.\ Phys.\  {\bf 45}, 3399 (2004);
Phys.\ Rev.\  D {\bf 74}, 125001 (2006);
Rev.\ Mex.\ F\'{\i}s. {\bf 56 (6)}, 469 (2010);
A.J.~Hariton and R.~Lehnert,
Phys.\ Lett.\  A {\bf 367}, 11 (2007);
D.~Colladay and P.~McDonald,
Phys.\ Rev.\ D {\bf 77}, 085006 (2008);
Phys.\ Rev.\ D {\bf D79}, 125019 (2009).

\bibitem{HigherDim}
R.C.~Myers and M.~Pospelov,
Phys.\ Rev.\ Lett.\  {\bf 90}, 211601 (2003);
C.M.~Reyes {\em et al.}, 
Phys.\ Rev.\ D {\bf 78}, 125011 (2008);
V.A.~Kosteleck\'y and M.~Mewes,
Phys.\ Rev.\ D {\bf 80}, 015020 (2009);
Phys.\ Rev.\ D {\bf 85}, 096005 (2012);
M.~Cambiaso {\em et al.}, 
Phys.\ Rev.\ D {\bf 85}, 085023 (2012);
G.~Rubtsov {\em et al.}, 
Phys.\ Rev.\ D {\bf 86}, 085012 (2012).

\bibitem{UHECR}
See, e.g., 
V.A.\ Kosteleck\'y and M.\ Mewes,
Phys.\ Rev.\ Lett.\  {\bf 87}, 251304 (2001);
Phys.\ Rev.\  D {\bf 66}, 056005 (2002);
R.~Lehnert and R.~Potting, 
Phys.\ Rev.\ Lett.\  {\bf 93}, 110402 (2004);
Phys.\ Rev.\  D {\bf 70}, 125010 (2004);
B.D.~Altschul,
Astropart.\ Phys.\  {\bf 28}, 380 (2007);
Phys.\ Rev.\  D {\bf 78}, 085018 (2008).

\bibitem{collider}
See, e.g., 
M.A.~Hohensee {\it et al.},  
Phys.\ Rev.\ Lett.\  {\bf 102}, 170402 (2009);
Phys.\ Rev.\  D {\bf 80}, 036010 (2009);
G.~Amelino-Camelia {\it et al.},
Eur.\ Phys.\ J.\  C {\bf 68}, 619 (2010);
V.~Gharibyan,
Phys.\ Lett.\ B {\bf 611}, 231 (2005);
J.-P.~Bocquet {\it et al.}
[GRAAL Collaboration],
Phys.\ Rev.\ Lett.\  {\bf 104}, 241601 (2010);
B.D.\ Altschul, 
Phys.\ Rev.\ D {\bf 80}, 091901 (2009);
Phys.\ Rev.\ D {\bf 84}, 076006 (2011).

\bibitem{cavities}
H.~M\"uller {\it et al.},
Phys.\ Rev.\ D {\bf 68}, 116006 (2003);
Ch.~Eisele {\it et al.},
Phys.\ Rev.\ Lett.\  {\bf 103}, 090401 (2009);
M.E.~Tobar {\it et al.}, 
Phys.\ Rev.\  D {\bf 80}, 125024 (2009).

\bibitem{neutrinos}
T.~Katori {\it et al.},
Phys.\ Rev.\ D {\bf 74}, 105009 (2006);
J.S.~D\'{\i}az {\it et al.},
Phys.\ Rev.\ D {\bf 80}, 076007 (2009);
V.~Barger {\it et al.},
Phys.\ Rev.\ D {\bf 84}, 056014 (2011); 
J.S.~D\'{\i}az and V.A.~Kosteleck\'y,
Phys.\ Rev.\ D {\bf 85}, 016013 (2012).

\bibitem{spectroscopy}
R.~Bluhm {\it et al.},
Phys.\ Rev.\ D {\bf 57}, 3932 (1998);
H.\ Dehmelt {\it et al.},
Phys.\ Rev.\ Lett.\ {\bf 83}, 4694 (1999);
R.\ Mittleman {\it et al.},
Phys.\ Rev.\ Lett.\ {\bf 83}, 2116 (1999);
G.\ Gabrielse {\it et al.},
Phys.\ Rev.\ Lett.\ {\bf 82}, 3198 (1999);
D.~Bear {\it et al.},
Phys.\ Rev.\ Lett.\  {\bf 85}, 5038 (2000);
R.~Lehnert and I.~Shimamura,
J.\ Plasma Fusion Res.\  {\bf 80}, 1006 (2004);
P.\ Wolf {\it et al.}, 
Phys.\ Rev.\ Lett.\  {\bf 96}, 060801 (2006);
I.~Altarev {\it et al.},
Phys.\ Rev.\ Lett.\  {\bf 103}, 081602 (2009);
Europhys.\ Lett.\  {\bf 92}, 51001 (2010).

\bibitem{AntiH}
See, e.g., R.~Bluhm {\it et al.},
Phys.\ Rev.\ Lett.\  {\bf 82}, 2254 (1999).

\bibitem{muonium}
See, e.g., V.W.~Hughes {\it et al.},
Phys.\ Rev.\ Lett.\  {\bf 87}, 111804 (2001).

\bibitem{annual}
See, e.g., F.~Can\`e {\it et al.},
Phys.\ Rev.\ Lett.\  {\bf 93}, 230801 (2004).

\bibitem{muon} 
See, e.g., 
G.W.~Bennett {\it et al.}  [Muon (g-2) Collaboration],
Phys.\ Rev.\ Lett.\  {\bf 100}, 091602 (2008).

\bibitem{mueller}
S.E.~M\"uller,
Hyperfine Interact.\  {\bf 215}, no.\ 1--3, 31 (2013).

\end{thebibliography}
\end{document}